\documentstyle[aps,prl,multicol,epsf,epsfig]{revtex}

\def\CC{{\rm\kern.24em \vrule width.04em height1.46ex depth-.07ex\kern-.30em
    C}}
\def\P{{\rm I\kern-.25em P}}
\def\RR{{\rm         \vrule width.04em height1.58ex depth-.0ex
    \kern-.04em R}}
\def\RR{{\rm\kern.24em \vrule width.04em height1.46ex depth-.07ex\kern-.30em
    R}}
\def\P{{\rm I\kern-.25em P}}
\def\RR{{\rm         \vrule width.04em height1.58ex depth-.0ex
    \kern-.04em R}}

\newcommand{\be}{\begin{equation}}
\newcommand{\ee}{\end{equation}}
\newcommand{\bq}{\begin{eqnarray}}
\newcommand{\eq}{\end{eqnarray}}

\newcommand{\al}{\alpha}
\newcommand{\bi}{\beta}

\newcommand{\ep}{\epsilon}

\newcommand{\ran}{\rangle}

\newcommand{\arccosh}{\text{arccosh}}

\begin{document}


\title{Optical Holonomic Quantum Computation\footnote{ICQI 2001 - International Conference on Quantum Information, Rochester, June 10-13, 2001 (Contribution given by D. E)}}

\author{Demosthenes Ellinas$^{1}$ and Jiannis Pachos$^{2}$}

\address{ 
$^1$Technical Univeristy of Crete, Department of Sciences, Section of Mathematics,
 GR-73 100 Chania Crete Greece, ellinas@science.tuc.gr\\
$^2$Max Planck Institut f\"{u}r Quantenoptik   
D-85748 Garching, Germany, jip@mpq.mpg.de
}

\maketitle

\begin{abstract}
Holonomic quantum computation (HQC) is materialized here with quantum optics
components. Holonomies are the generalization of the Berry phases to unitary matrices 
with dimensionality the same as the degree of degeneracy of the system. In a nonlinear Kerr 
medium the degenerate states of laser beams are interpreted as qubits. Control 
manipulations with displacers, squeezers and two-mode interfering devices performed in a cyclic, 
adiabatic fashion produce holonomies. Here, they are employed as logical gates for our 
HQC proposal. The effects of errors from imperfect control of classical parameters, the 
looping variation of which builds up holonomic gates, are investigated.

\end{abstract}

\date{today}

\vspace{1cm}
\copyright 1999 Optical Society of America

{\bf OCIS codes}:(999.9999) Quantum Computing

\vspace{0.5cm}

Quantum computers consist of a tensor product structure of two level quantum sub-systems (qubits)
which are manipulated coherently with a large set of external operations (logical gates). Here
an application of holonomic manipulations \cite{zanardi} (see also \cite{jones}) with optical
components of laser Fock states is
described. Holonomy is the generalization of the well known Berry phase to the case of a higher
dimensional unitary matrix \cite{pachos}. It is produced by an adiabatic cyclic evolution of external control
parameters that drive a degenerate Hamiltonian\cite{shapere}. This Hamiltonian in our case is produced by a
Kerr medium through which the lasers are passing \cite{pachos1}. Hence, a two fold degeneracy is created, where
each qubit is encoded. As the first part of this presentation shows the logical gates are constructed
by employing displacings and squeezings of each mode for producing one-qubit gates. Two-qubit gates
are produced by two mode interfering devices like two-mode displacing or squeezing. Each laser encodes
one qubit in its degenerate sub-space, spanned by the Fock states $|0\ran$ and $|1\ran$. By alternative
manipulations with the displacing operator $D(\lambda)=\exp ( \lambda a^\dagger  - \bar \lambda a)$ and the 
squeezing operator
$S( \mu )= \exp ( \mu \left. a^\dagger \right.^2  - \bar \mu a^2 )$ in an adiabatic fashion it is possible to 
generate any one-qubit rotation.
Adiabaticity along the cyclic evolution of the parameters $\lambda=x+iy$ and $\mu=r_1 \exp i \theta_1$ 
guarantees that the states $|0\ran$ and $|1\ran$ will not mix with
the rest of the Fock states of  the laser. On the other hand as they have the same energy they are
allowed to exchange population along this cyclic evolution. 
The exact evolution they undertake at the end of this manipulation is given by unitaries which depend
on the geometrical characteristics of the cyclic procedure \cite{pachos}. The
coordinates of the space where the loops belong are not abstract theoretical
objects but rather the parameters the experimentalist is tuning in the lab,
like the squeezing or the displacing amplitude. For a loop $C_I$ on the
$(x,r_1)_{\theta_1=0}$ plane we obtain the holonomy 
\be
\Gamma_A ( C_I )= \exp (-i \sigma_1 \Sigma_I ) \, \, ,
\ee
where 
$\Sigma_I :=\int_{D(C_I)}dx dr_1 2 e^{-2r_1}$ . Hence it represents the area on a hyperbolic
surface enclosed by the loop $C_I$. For this procedure displacings and squeezings are alternated. 
Similarly we obtain for a loop $C_{II}$ on $(x,r_1 )_{\theta_1=\pi/2}$ the holonomy
\be
\Gamma(C_{II} )= \exp (-i \sigma_2 \Sigma_{II} )
\ee
with $\Sigma_{II} :=\int _{D(C_{II})} 2 e^{-2 r_1}$ .

For the case of two-qubit gates we use two laser beams passing through a Kerr medium. This produces
a 4-dimensional degenerate space with a tensor product structure. Two-mode interfering devices produce 
two-mode displacing  transformation $N(\xi)=\exp( \xi a_1^\dagger a_2  - \bar \xi a_1  a_2 ^\dagger )$ 
or two-mode squeezing $M(\zeta)=\exp( \zeta a_1 ^\dagger a_2^\dagger  - \bar \zeta a_1 a_2 )$. 
Interchanging these procedures in a cyclic, adiabatic fashion produces two-qubit unitaries. In particular, 
for $\zeta=r_2 e^{i\theta_2}$ and $\xi=r_3 e^{i \theta_3}$ we have a four-parameter real manifold. 
Taking a loop $C_{III} \in  (r_2 ,r_3 )_{\theta_2=\theta_3=0}$ we obtain 
\be
\Gamma ( C_{III} )= \exp (-i \sigma^{12}\Sigma_{III}  )
\ee
with $\Sigma_{III} :=\int_{D(C_{III})} dr_2 d r_3 2 \sinh (2 r_2)$. This is a
nontrivial unitary transformation 
and hence, together with the general one-qubit holonomic transformations they
can produce any desired logical  
gate resulting to universality. Moreover, the ability to produce a big variety
of gates by simply changing the experimental parameters and spanning different
loops reduces the amount of resources needed to construct certain gates in the
traditional way by the successive combination of basic gates. 

In the previously proposed setup the degeneracy has to be present at all times. 
To achieve this along with the implementation of the adiabatic transformations with the control 
devices it is necessary to employ the ``kicking" method. The laser beams are inserted into the 
nonlinear Kerr medium for time $\Delta t$ and alternated with infinitesimal transformation 
$D(\delta \lambda )$, $S(\delta\mu )$, $M(\delta\zeta )$ or $N(\delta \xi )$ which take place 
in times $\delta t  \ll \Delta t$. This ensures the presence of the degeneracy along the adiabatic evolution. 

Holonomic quantum computation is an all geometric setup of quantum evolution. The parameters of the gates are 
areas of particular surfaces. Deformations of these contours due to statistical experimental errors does not 
change the value of the areas to the first order of  approximation. Moreover, the weight factors in the 
parameters  $\Sigma_I$ and $\Sigma_{II}$ 
of the corresponding gates are exponentially decreasing for large $r_1$. This makes the gates resilient 
to displacing control errors as they are suppressed exponentially due to squeezing \cite{pachos1}. 
Such characteristics are important for error avoiding computation. 

Next we aim to discuss the error avoiding features of the optical holonomic proposal in the presence 
of systematic errors.
More specifically the parametric loops producing holonomies, 
when subject to imperfections while spanned, introduce an error in the final gates through 
their accordingly fluctuated parameters. This is systematically studied for the Hadamard (H) 
gate and a non trivial two qubit gate. 
In order to study the errors introduced by imperfect control of the external parameters we 
adopt an imperfectly spanned loop, $C^\prime$ \cite{ellinas}. This by its turn
will produce an area as a gate parameter  
given by $\Sigma^\prime =\Sigma +\ep$. If the errors are statistical rather than systematical then 
the area spanned by this loop are, to the first order, zero. Let us consider how systematic errors 
in the area effect one and two qubit gates. The Hadamard gate is given by 
\be
 \begin{array}{cc}
U_H=&  
 \left[  \begin{array}{ccc}    \cos \Sigma & \sin \Sigma \\
                \sin \Sigma & -\cos \Sigma \\
\end{array} \right] \,\, , 
\end{array}
\ee
for $\Sigma=\pi/4$. Up to a corrective phase it may be produced by a loop $C_{II}$, with spanning area 
given by $\Sigma =\int_{D(C_{II})} d x d r_1 2 e^{-2 r_1}$ with $D(C_{II})$
taken to be a rectangular surface enclosed  
by $\{0 \leq x \leq \pi /4 , 0\leq r_1 \leq \ln 2\}$. Introduce an error in this surface by translating 
the boarders of $x$ and $r_1$ by $\al$ and $\bi$ respectively, where $\al,\bi \ll 1$. This is a kind of 
systematic error. The imperfect Hadamard gate is given to the first order in $\ep$ by $U_H (\ep)=U_H+\ep h$, with 
$\begin{array}{cc}
h=&{1 \over \sqrt{2}}   
\left[  \begin{array}{ccc}   
                -1 & 1 \\  
                 1 & 1 \\
\end{array} \right]
\end{array}$.
For large values of $r_1$ the dependence of $\ep$ on $\beta$ is exponentially
small, while the error of $x$  
is introduced linearly, i.e. $\ep=\alpha$. As the squeezing of the light beam 
is much harder to perform and to 
control in experiments compared to its displacing the above feature is very
appealing. Theoretically this characteristic makes the passage from the
geometrical to the topological quantum computation as $\beta$ errors in the
spanning of the loop $C^\prime$ are suppressed even if they are very large\cite{pachos2}.

For a two qubit gate we perform a loop $C_{III}$ between the variables $r_2$ and $r_3$ with area 
$\Sigma_{III}={\pi \over 4}$ which eventually gives the gate
\be
\begin{array}{cc}U=& {1\over \sqrt{2}}   
\left[  \begin{array}{cccc}   
           \sqrt{2} & 0 & 0 & 0 \\
           0 & 1 & -1 & 0 \\ 
           0 & 1 &  1 & 0 \\
           0 & 0 &  0 & \sqrt{2} \\
\end{array} \right]
\end{array}
\ee
Allowing the area $\Sigma_{II}$ to be enclosed by the following rectangular 
$\{0 \leq r_2 \leq \arccosh(2) , 0\leq r_3 \leq \pi /8 \}$ then systematic small errors in 
the definition of the boarders of the rectangular of the form $(\alpha',\beta')$ gives an error in the 
gate parameter of the form $\Sigma'=\Sigma +\delta$ with $\delta= 1.7 \alpha'+ \beta'$. The gate in this 
case is given by $U(\delta)=U+\delta u$ where
\be
\begin{array}{cc}
u=& {1\over \sqrt{2}}   
\left[  \begin{array}{cccc}   
 0 & 0 &  0 & 0 \\
 0 & -1 & -1 & 0 \\ 
 0 & 1 &  -1 & 0 \\
 0 & 0 &  0 & 0 \\
\end{array} \right]
\end{array}
\ee
We notice that with the ordering $\{|00 \ran, |10\ran, |11\ran, |01\ran\}$, adopted for the basis vectors that span the
four dimensional degenerate subspace of coding the quantum information, where the first qubit is the control and 
the second is the target, the $U$ gate is a control $\pi/4$-rotation. Introducing the control phase gate 
$P_\phi=diag(1,e^{i\phi},1,1)$, we can construct the control-not gate with a control sign flip operation as 
$U_{CN}=P_\pi U^2$. Then the holonomic control not gate subjected to systematic errors is 
$U_{CN}(\delta)=U_{CN} + \delta P_\pi u$.

By means of the imperfect one and two qubit quantum gates so obtained by the optical model
of holonomic computation \cite{pachos1}, we can now proceed along the lines of \cite{ellinas}, 
to provide an all optical physical implementation of the holonomic teleportation \cite{gottesman,nielsen} 
and the holonomic remote gate construction \cite{shor,zhou}, that have been put forward in \cite{ellinas},
in the more mathematical framework of the $CP^n$ model.

Finally we should mention that alternatives to the optical implementation of holonomies for quantum computation
have been suggested recently \cite{duan}, in a scheme that uses adiabatic manipulations of parameters 
determining the interaction between laser beams and trapped ions in order to generate abelian and non-abelian
geometrical gates. In such a model in addition to the systematic control errors presented here one has to 
take in account other imperfection mechanisms such as interaction with the environment and the effect of the
spontaneous atomic emission upon Berry connection, see e.g. \cite{ellinas1}.



\begin{references}

\bibitem{zanardi} P. Zanardi and M. Rasetti, Phys. Lett. A {\bf 264}, 94 (1999).

\bibitem{jones} J. Jones, V. Vedral, A. Ekert and G. Castagnoli, Nature {\bf 403}, 869 (1999).

\bibitem{pachos} J. Pachos, P. Zanardi and M. Rasetti, Phys. Rev. A {\bf 61} 010305 (R) (2000).

\bibitem{shapere} A. Shapere and F. Wilczek, ``Geometric Phases in Physics", (World Scientific 1989).

\bibitem{pachos1} J. Pachos and S. Chountasis, Phys. Rev. A {\bf 62}, 052318 (2000).

\bibitem{ellinas} D. Ellinas and J. Pachos, ``Universal Holonomic Quantum   Computing and Imperfections'', 
quant-ph/0009043, revised version to appear in Phys. Rev. A.

\bibitem{pachos2} J. Pachos, G. Morigi, W. Lange and H. Walther, to appear.

\bibitem{gottesman} D. Gottesman, in Proc. of the XXII Inter. Symp. on Group
  Theor. Methods in Physics, eds. S. P. Corney, R. Delbourgo and P. D. Jarvis,
  p. 32 (Cambridge, MA International Press), quant-ph/9807006.

\bibitem{nielsen} M. A. Nielsen and I. L. Chuang, ``Quantum Computation and
  Quantum Information", (Cambridge Univ. Press, 2000).

\bibitem{shor} P. W. Shor, Proc. 35th Annual Symp. on Fundamentals of Computer
  Science (IEEE Press, Los Alamos, 1996) p. 56, quant-ph/9605011).

\bibitem{zhou} X. Zhou, D. W. Leung and I. L. Chuang. ``Quantum Logic Gate
  Constructions with one-bit ``teleportation"", Phys. Rev. A {\bf 62}, 052316
  (2000), quant-ph/0002039.

\bibitem{duan} L.-M. Duan, J. I. Cirac, and P. Zoller, Science 2001 June 1;
  292: 1695.

\bibitem{ellinas1} D. Ellinas, S. M. Barnett and M.-A. Dupertuis, Phys. Rev. A
  {\bf 39} 3228 (1989).


\end{references}
\end{document}